\documentclass[
preprint,
 amsmath,amssymb,
 aps,
]{revtex4-1}

\usepackage{graphicx}
\usepackage{dcolumn}
\usepackage{bm}

\usepackage{comment}
\usepackage{float}
\usepackage{amsmath}
\usepackage{cases}
\usepackage{listings}
\usepackage{xcolor}
\usepackage{subfigure}
\usepackage[colorlinks]{hyperref}

\begin{document}

\title{Trajectory Prediction of Rotating Objects in Viscous Fluid: Based on Kinematic Investigation of Magnus Glider}
 \author{Zhiyuan Wei${}^1$}%
 \author{Lijie Ding${}^2$}
  \author{Kai Wei${}^2$}
   \author{Ziwei Wang${}^2$}
  \author{Rucheng Dai${}^3$}
  \email{dairc@ustc.edu.cn}
\affiliation{${}^1$School of Gifted Young, University of Science and Technology of China, Hefei 230026, Anhui Province, China}
 \affiliation{${}^2$School of Physics, University of Science and Technology of China, Hefei 230026, Anhui Province, China}
 \affiliation{${}^3$The Centre of Physical Experiments, University of Science and Technology of China, Hefei 230026, Anhui Province, China}



\begin{abstract}

The case of a rotating object traveling through viscous fluid appears in many phenomena like the banana ball and missile movement. In this work, we build a model to predict the trajectory of such rotating objects with near-cylinder geometry. The analytical expression of Magnus force is given and a wind tunnel experiment is carried out, which shows the Magnus force is well proportional to the product of angular velocity and centroid velocity. The trajectory prediction is consistent with the trajectory record experiment of Magnus glider, which implies the validity and robustness of this model.

\end{abstract}

\maketitle

\section{Introduction}
Magnus force was first described by G. T. Walker in 1671 when observing tennis ball \cite{benedetti1989flight}. In 1742, B. Robins explained the deviation of the trajectories of musket balls by the Magnus effect \cite{lorenz2011planetary,steele1994muskets}. Magnus force occurs when rotating objects travel through the air with an angle between the axis of rotation and the flight velocity\cite{briggs1959effect,brown1971see,van1982album}. Magnus force is perpendicular to the rotating axis and the flight direction, while its magnitude is decided by the movement and geometry of the body and properties of the fluid such as viscous coefficient and density. H. G. Magnus ascribed the asymmetrical transverse force to the pressure difference performing on the surface of the object produced by Bernoulli effect because the rotation brings additional velocity at the edge \cite{magnus1852uber,magnus1853ueber}.

The Magnus effect changes the trajectory of rotating objects in the air, which is important in many fields like sports competition and ballistic problems \cite{clancy1975aerodynamics}. In ballistics, people usually force the shell and missile spinning to stabilize them. Thus it is essential to involve Magnus force. In sports, Magnus effect causes some dramatic movement such as the banana ball and slice. Although this phenomenon has been noticed and studied for a long time, precise predictions for particular processes can be hardly made, which is ascribed to the complexity of mathematical formulation and solution \cite{swanson1961magnus,mittal2003flow,thouault2012numerical,seifert2012review}. Nevertheless, in some simple cases the Magnus effect of rotating objects was investigated theoretically and experimentally, like the two-dimensional cylinder\cite{benedetti1989flight,swanson1961magnus,seifert2012review} and three-dimensional sphere\cite{briggs1959effect,hess1968coupled,watts1987lateral}.

The trajectory prediction of spinning object is hardly reported. In this paper, we build a trajectory prediction model based on the kinematic behavior of Magnus glider, then test it with a wind tunnel. 

The rest of the paper is organized as follows. In Sec. II we theoretically analyze the origin of Magnus force and develop a model to give the equation of motion for Magnus glider under approximation. In Sec. III, a wind tunnel is built to measure the Magnus force, and the trajectory of Magnus glider is recorded. Furthermore we give trajectory prediction and compare it with the experiment result. Finally we conclude in Sec. IV.

\section{Theoretical Model for Magnus Glider}

    \subsection{Origin of Magnus Force}

\begin{figure}[!h]
          \centering
          \includegraphics[width = 0.7\textwidth]{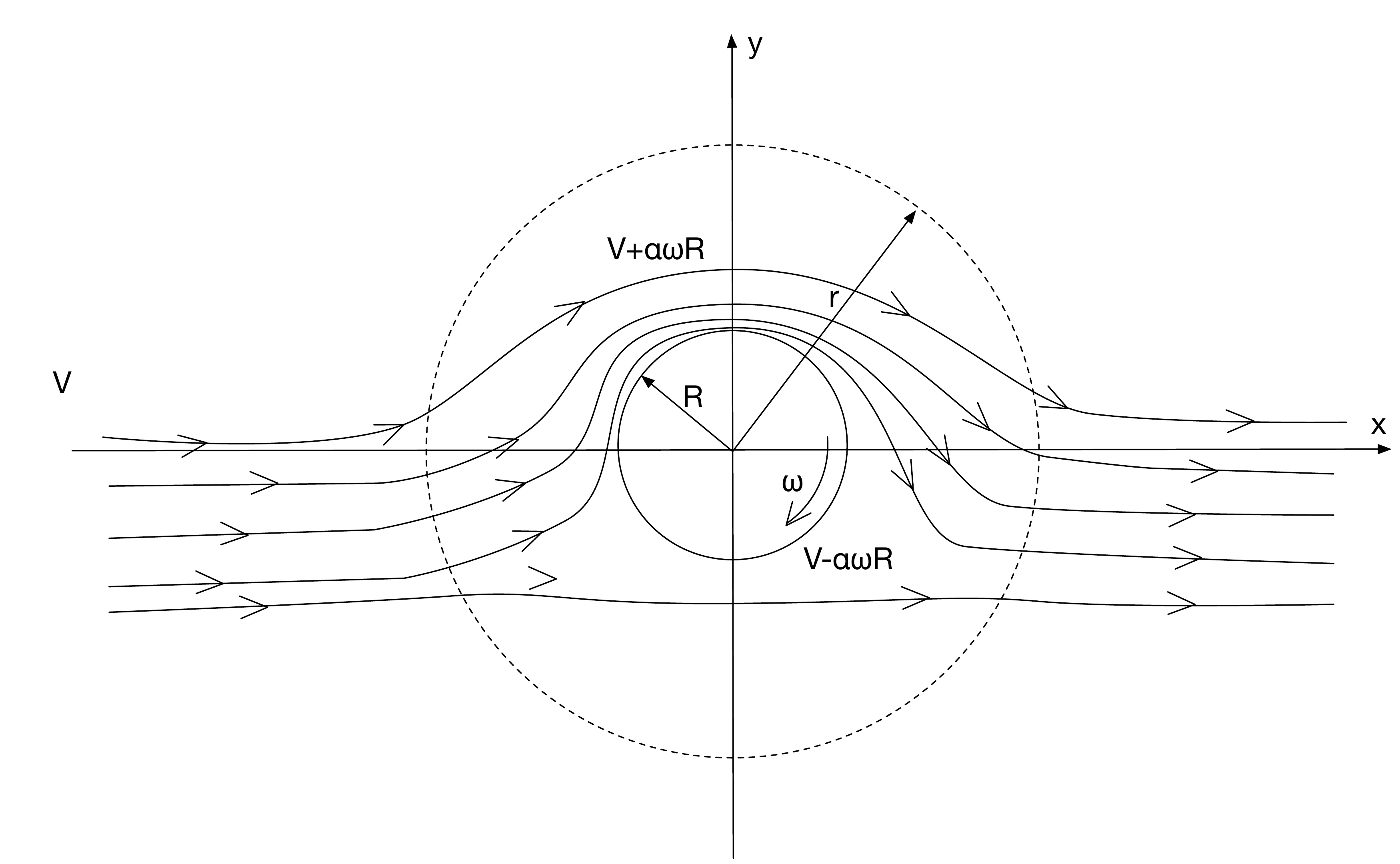}
          \caption{Schematic graph of fluid field around cylinder.}
          \label{fig: Fluent around cylinder}
        \end{figure}

        In this section, we analyze the origin of Magnus Force. Considering the system in the instantaneous inertial frame of the glider. In this case, the glider is rotating while fluid flowing around it, shown in Fig.~\ref{fig: Fluent around cylinder}.

        Consider a long cylinder ($L \gg R$) rotating in a perfect fluid with angular velocity $\omega$ and centroid velocity $V$. In the center-of-mass frame of the cylinder, the cylinder will only have angular velocity in the fluid flow. The equation of fluid dynamics can be solved under irrotational and incompressible condition. For a long cylinder, translation symmetry along axes $z$ is approximately preserved which simplify the equations to two-dimensional case.

        The velocity potential $\phi \left( {r,\theta } \right)$ satisfy the Laplacian Equation~\eqref{Lap}.

        \begin{equation}
        \label{Lap}
          \frac{1}{r}\frac{\partial}{\partial r}(r\frac{\partial\phi}{\partial r}) + \frac{1}{r^2}\frac{\partial^2\phi}{\partial \theta ^2} = 0
        \end{equation}

        We denote the circulation around the cylinder as $\Gamma$, also consider the $r\to\infty$ asymptotic case, we have ~\eqref{Bon}

        \begin{equation}
        \label{Bon}
          \lim_{r\rightarrow\infty}\phi(r,\theta ) = -V r\cos\theta + C~~,~~\lim_{r\rightarrow R}\oint \vec{v}\cdot d\vec{l} = \lim_{r\rightarrow R}\oint \frac{\partial\phi}{\partial\theta}d\theta = \Gamma
        \end{equation}

        The solution of $\phi(r,\theta)$~\eqref{Eul} can be obtained from~\eqref{Lap} and~\eqref{Bon}.

        \begin{equation}
            \label{pot}
            \phi(r,\theta) = V(1+\frac{R}{r^2})r\cos\theta - \frac{\Gamma}{2\pi}\theta
        \end{equation}

        Magnus force, which is the projection of fluid force on the y direction, can be calculated from velocity potential. In the $x$ direction of Fig.~\ref{fig: Fluent around cylinder} , centroid speed on both sides are the same. Magnus force is given by Euler formula~\eqref{Eul}:

        \begin{equation}
            \label{Eul}
            \int_\tau \frac{\partial}{\partial t}(\rho V)d\tau + \int_s \rho v_r V ds = -\int_s p \sin\theta ds + \int_\tau \rho f_y d\tau -F_{Mag.}
        \end{equation}

        The first term of left hand side of ~\eqref{Eul} can be ignored since the fluid field is a approximately static field in our situation. Besides, there is no momentum transfer across the surface $r = R$ since the fluid cannot enter the cylinder. By substituting ${v_r} = {\partial \phi }/{\partial r} = {V}\left( {1 - \frac{{{R^2}}}{{{r^2}}}} \right)\cos \theta $ and ${v_\theta }=\left(1 + {R^2}/{r^2}\right)\sin \theta  - \Gamma/{2\pi r}$ to \eqref{Eul}, Magnus force on y direction can be simplified as~\eqref{Mfp}.

        \begin{equation}
            \label{Mfp}
            \begin{aligned}
            F_{mag.} & = -\rho V \Gamma L & = - 2\pi \alpha R^2\omega\rho V L
            \end{aligned}
        \end{equation}

        \eqref{Mfp} shows that Magnus force is proportional to the circulation of fluid, which depends on a lot of conditions such as material properties and geometry of Magnus glider. 
        
        Now we introduce a phenomenal simplification to conduct analytical results: Suppose the change of fluid velocity near the cylinder surface is proportional to rotating velocity of the cylinder with coefficient $\alpha$, thus the velocity change is $\alpha\cdot \omega R $ (See in Fig.~\ref{fig: Fluent around cylinder}). Substitute this to \eqref{Bon} and we get the circulation $\Gamma = 2\pi \alpha R^2\omega$.

    \subsection{Equation of Motion for Magnus Glider }

        \begin{figure}[!h]
          \centering
          \includegraphics[width = 0.3\textwidth]{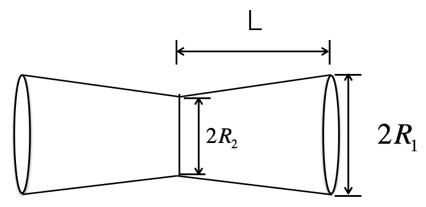}
          \caption{Geometry of Magnus glider, consisting two truncated-cone paper cups. $R_1$, $R_2$ denote the outer and inner radius of the cup, while $L$ denotes the height of the truncated cone.}
          \label{fig: Shape of Magnus Glider}
        \end{figure}

        The Magnus glider consists of two truncated-cone paper cups, shown in Fig.~\ref{fig: Shape of Magnus Glider}. Since it is not strictly a cylinder, we need to make a correction in our equation to reflect the geometry deviation. By integral over the surface (do not contain the inner side) of Magnus glider, the Magnus force~\eqref{eq:Magnus_force} is obtained.

        \begin{equation}
            \label{eq:Magnus_force}
            F_{Mag.} = 2\int_0^L {\Delta P \cdot r \cdot dl\int_0^\pi  {\sin \theta d\theta } }  = \frac{4}{3}\alpha \rho V\omega l\left( {R_1^2 + {R_1}{R_2} + R_2^2} \right)
        \end{equation}

        On the other hand, the fluid resistance generally has the form $\vec{f} =  - \gamma \vec{v}$, where $\gamma$ is the coefficient of resistance.

       The dynamic evolution of a Magnus glider can be predicted under different initial conditions since all the three dominant forces, Magnus, viscous and gravity force, are obtained.

\section{Experiment Test with Magnus glider}

    \subsection{Measurement of Magnus Force}

        \begin{figure}[!h]
            \centering
            \subfigure[Apparatus Setup]{\includegraphics[width = 0.3\textwidth]{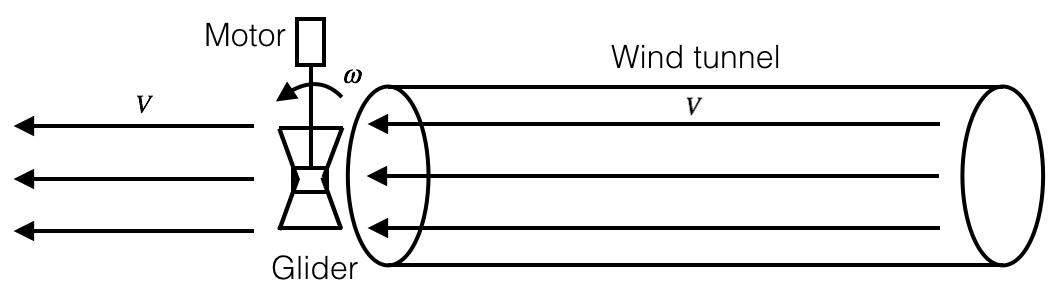}}
            \subfigure[Details for Spinning Cylinder]{\includegraphics[width = 0.3\textwidth]{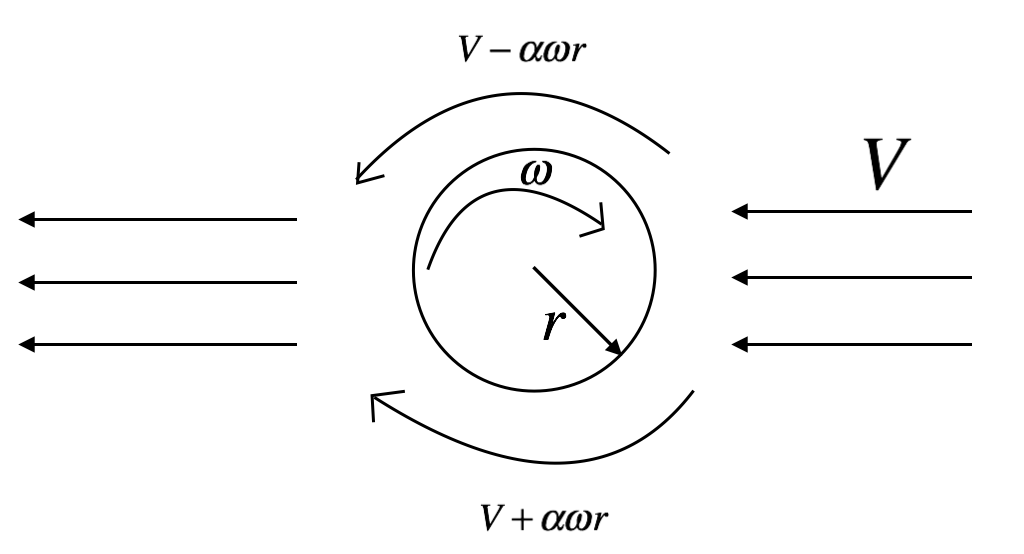}}
            \subfigure[Force Measurement]{\includegraphics[width = 0.2\textwidth]{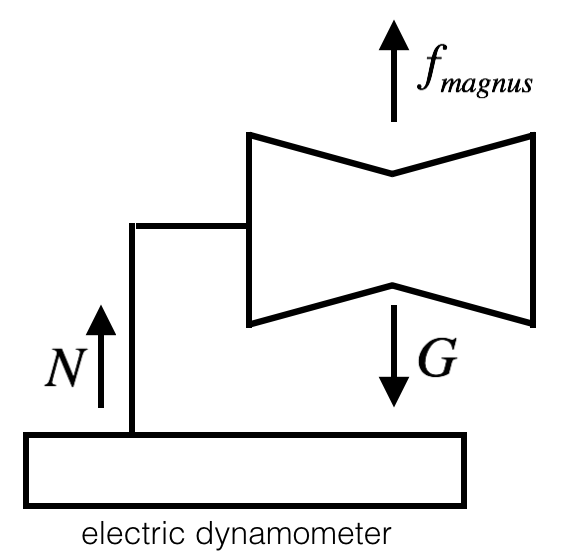}}
            \caption{Sketch of wind tunnel experiment. The length of the cylinder wind tunnel is $2m$, and the diameter of it’s open surface circle is $34cm$. The apparatus is placed on the steady ground. (a) shows the basic structure of wind tunnel experiment. We use an electric fan and a long plastic pipe to produce a wide range of stable wind field. The Magnus glider locates on the nozzle of the pipe and is linked with a motor to make it rotating along its central axis. By measuring the change of supportive force $N$ in (c), we can get the magnitude of Magnus force.}
            \label{fig: sketch wind}
        \end{figure}

\begin{figure}[!h]
          \centering
          \includegraphics[width = 0.65\textwidth]{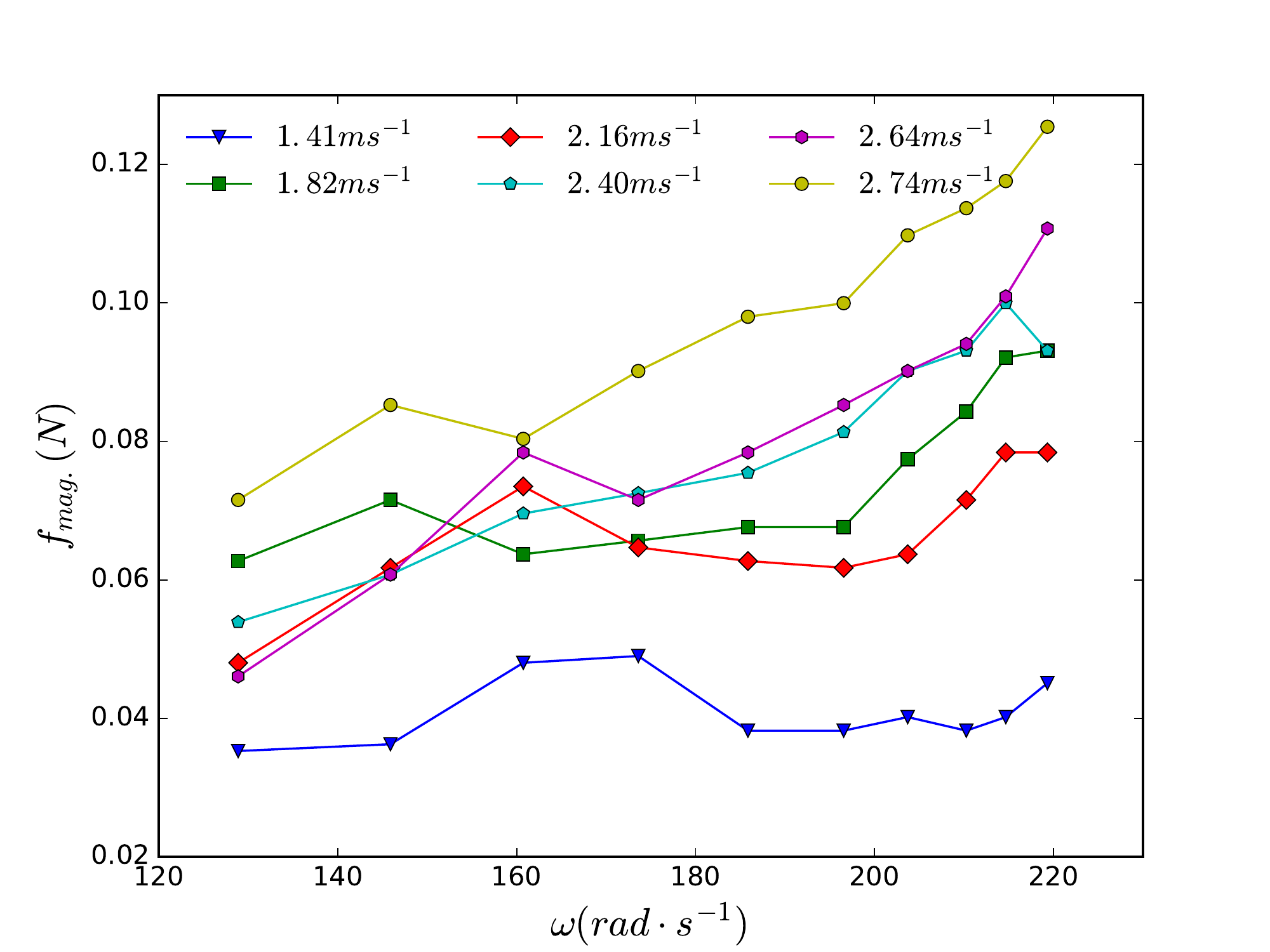}
          \caption{Magnitude of Magnus force vs rotation speed $\omega$ under different wind speed $V$. Different lines represent different wind speed through the tunnel.}
          \label{fig: Magnus force with rotation speed and wind speed}
        \end{figure}

\begin{figure}[!h]
          \centering
          \includegraphics[width = 0.65\textwidth]{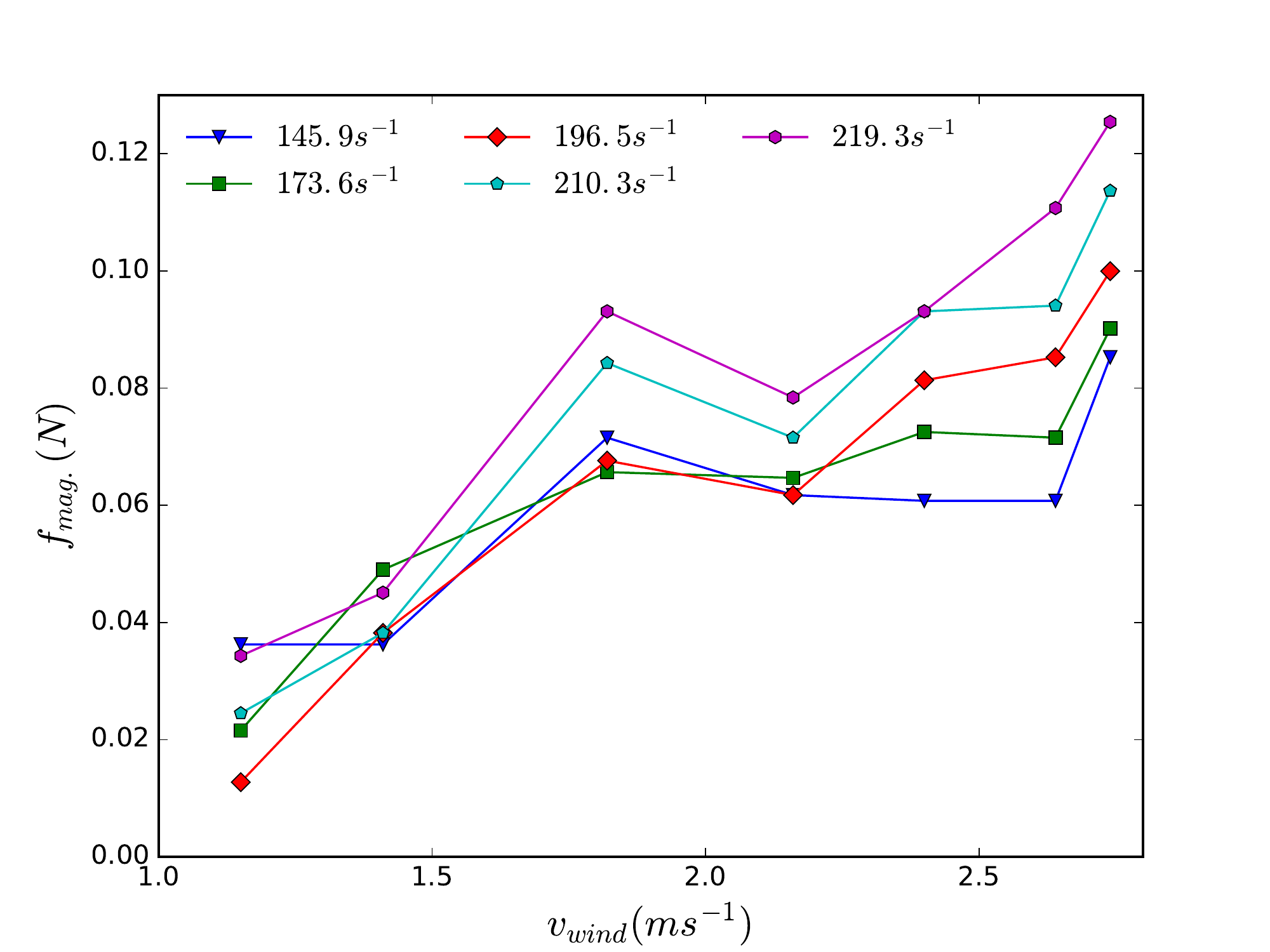}
          \caption{Magnitude of Magnus force vs wind speed $V$ under different rotating speed $\omega$. Different lines represent different angular velocity of spinning cup.}
          \label{fig: Magnus force with rotation speed and wind speed2}
        \end{figure}

        To figure out the relation between Magnus force and kinematic parameters, or more explicitly to get the value of drag coefficient α, a small wind tunnel was built (see Fig.~\ref{fig: sketch wind}). Relations between Magnus force $f_{Mag.}$ , rotation speed of Magnus glider $\omega$ and centroid speed $V$ were investigated and shown in Fig.~\ref{fig: Magnus force with rotation speed and wind speed} and Fig.~\ref{fig: Magnus force with rotation speed and wind speed2}.

\begin{figure}[!h]
            \centering
            \includegraphics[width = 0.7\textwidth]{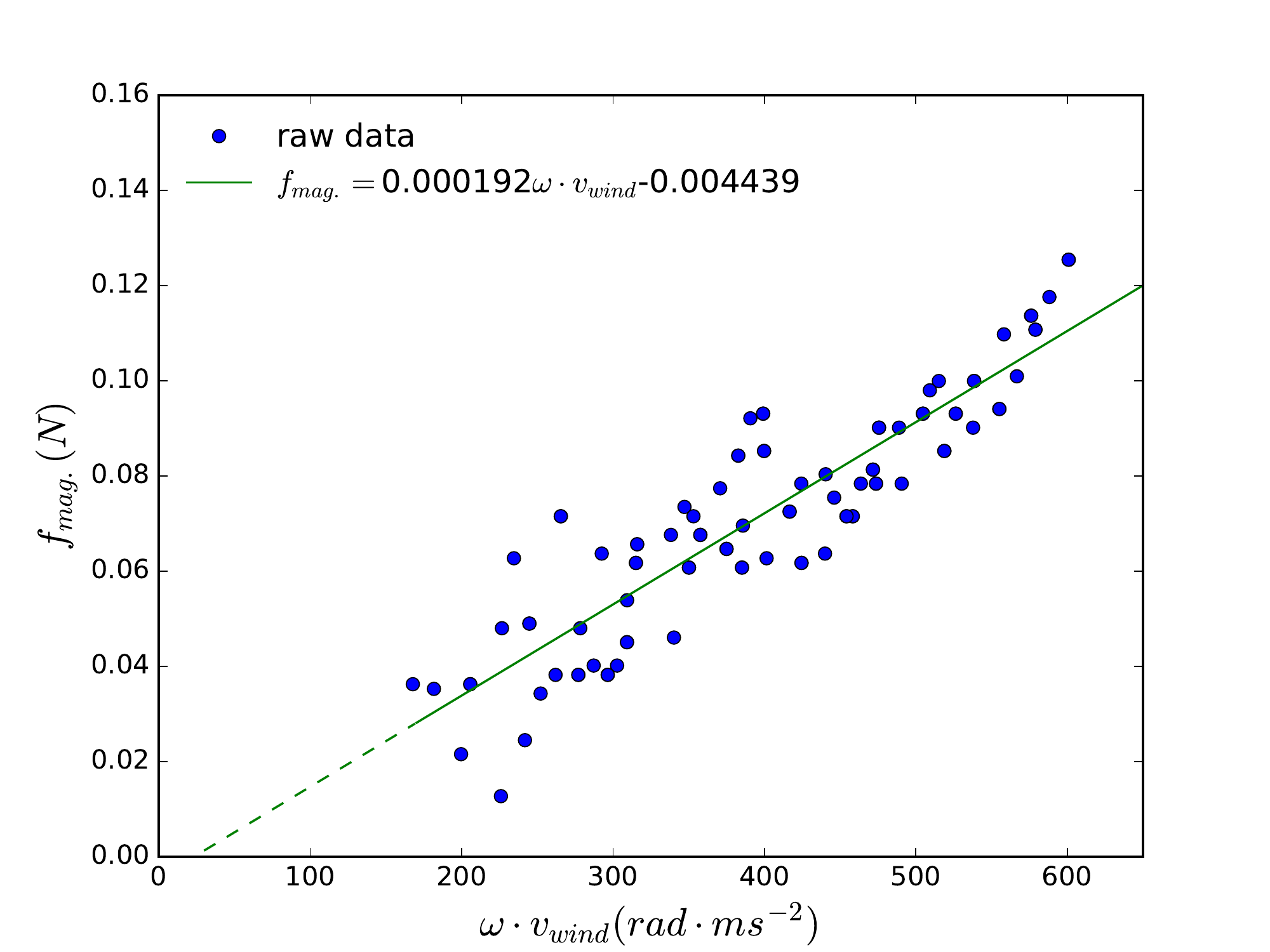}
            \caption{Magnitude of Magnus force vs the product of rotation speed $\omega$ and wind speed $V_f$ . The linear correlation coefficient $r = 0.957$, showing the $F_Mag.$ basically be proportional to ${V_f} \times \omega $.}
            \label{fig: Magnus force with product of rotation speed and wind speed}
        \end{figure}

        Fig.~\ref{fig: Magnus force with product of rotation speed and wind speed} shows the magnitude of Magnus force versus the product of rotation speed and wind speed. It shows a linear relation with the correlation coefficient $r = 0.957$, which is consistent with the theoretical model result~\eqref{Mfp}.

        Using data from the wind tunnel in Fig.~\ref{fig: Magnus force with product of rotation speed and wind speed}, we can get the parameter $\alpha$ by the formula \eqref{eq: get alpha}
        \begin{equation}
            \alpha  = \frac{{3F}}{{4\omega v\rho l\left( {R_1^2 + {R_1}{R_2} + R_2^2} \right)}}
            \label{eq: get alpha}
        \end{equation}

        Substituting $\rho  = 1.206kg/{m^3}$ and $g = 9.80m/{s^2}$, we get $\alpha = 0.1207$. From the experiment we notice the system only has slight dragging.

    \subsection{Capture for Motion of Magnus glider}

\begin{figure}[!h]
            \centering
            \includegraphics[width = 0.7\textwidth]{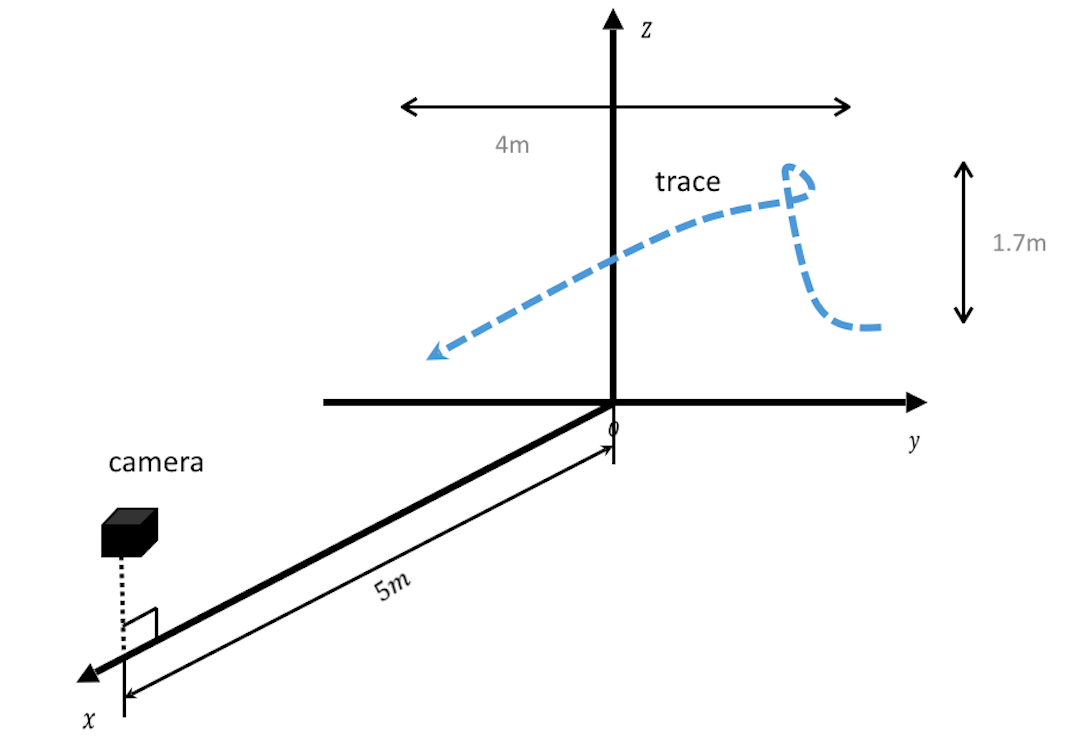}
            \caption{Magnus emission experiment: a video camera with $24fps$ (Frame per Second) is fixed to record the whole period of motion. Notice that the plane of trajectory of Magnus glider is parallel to the wall behind it.}
            \label{fig:Capture for Magnus Glider Motion}
        \end{figure}

        After we get some knowledge of origin of Magnus force, in principle it is not difficult to test our theory with the real motion of Magnus glider. To get the information of the movement of glider, we did the Magnus glider emission experiment in Fig.~\ref{fig:Capture for Magnus Glider Motion}. A fixed video camera with $24fps$ to record the motion, and from the video we can extract the position data of Magnus glider’s motion.

        The movement can be basically divided into three parts (see in Fig.~\ref{Trajectory}). In part I, the glider accelerates because of the tension of rubber band. In part II, the glider moves toward the highest point with deceleration. In part III, the glider moves uniformly before landing. From Fig.~\ref{Trajectory}, it can be clearly seen that the latest part of the movement of Magnus glider is a smooth uniform movement. In this part, the gravity, fluid resistance and Magnus force balance with each other.

    \subsection{Trajectory Prediction}

        \begin{figure}[!h]
            \centering
            \includegraphics[width = 0.7\textwidth]{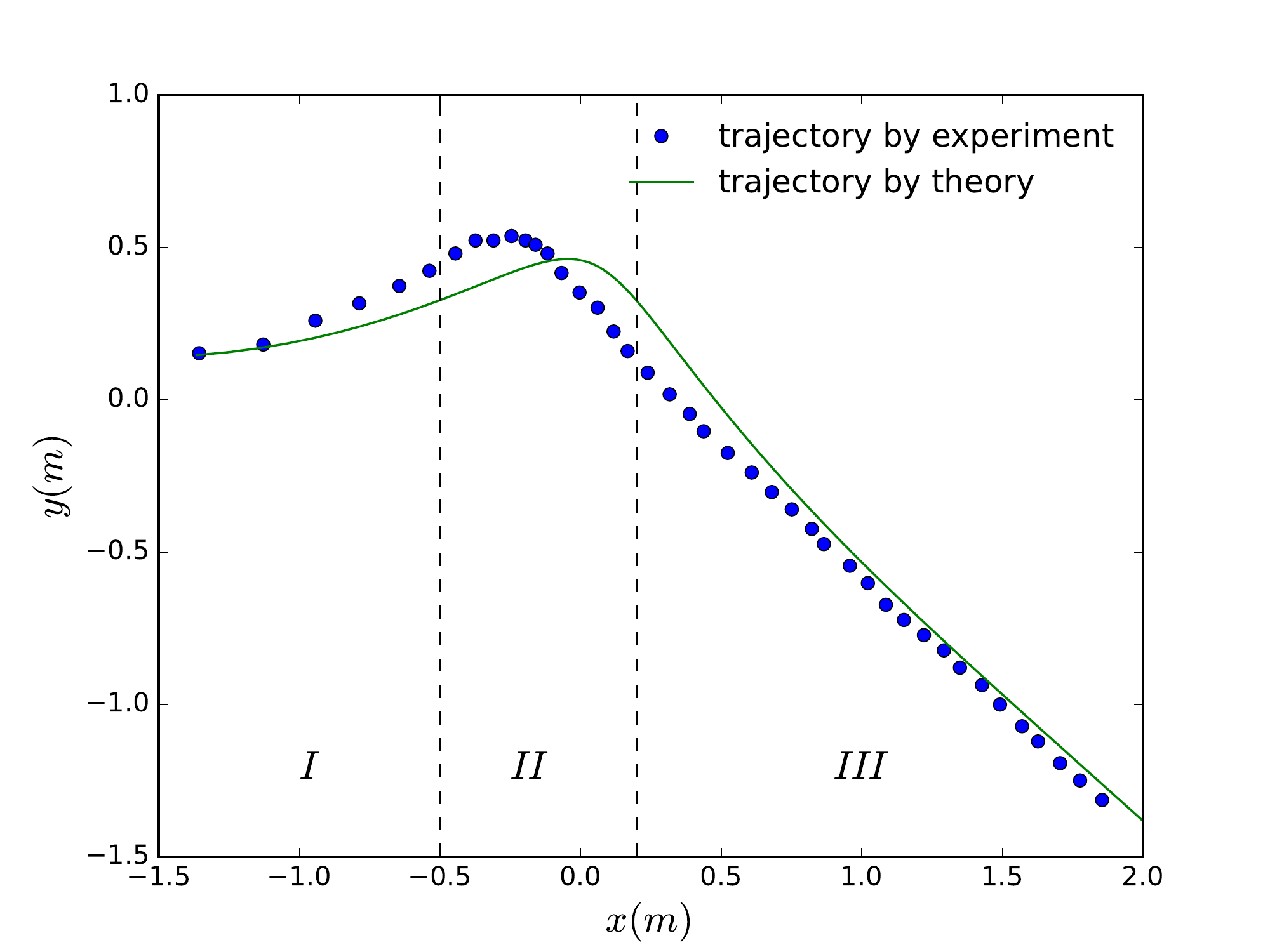}
            \caption{(Case 1) Trajectory of Magnus glider. Green line is the theoretical prediction by our model. The launch angle is ${5^ \circ }$ to the horizontal.}
            \label{Trajectory}
        \end{figure}

        Using our model, we can fully predict the evolution of the Magnus glider`s position. Here we show two cases to demonstrate the prediction ability of our theory. In case 1 (Fig.~\ref{Trajectory}), the parameters of the Magnus glider are $R_1 = 7.3cm$ , $R_2 = 5.2cm$, $L = 8.3cm$, $m = 9.1g$. The trajectory data is recorded by a high-speed video camera ($240fps$) . The rotation speed is $\omega  = 26.7 \cdot 2\pi~rad/s$, and the launch speed of the Magnus glider is ${V} = 2.01m/s$ (extracted from the motion video). The relative change of rotating speed in moving process is less that 10\%. We can calculate the Reynold number ${\mathop{\rm Re}\nolimits}  = \frac{{\rho VD}}{\mu }$ of the fluid around the surface of Magnus glider.

        Substitute the parameters into the expression : $\rho = 1.206kg/m^3$, $V = 2.01m/s$, $D = 15cm$ and $\mu = 3.52×10−5Pa\cdot s$, we have ${\mathop{\rm Re}\nolimits}  = 1.033 \times {10^4}$. This Re is already near the value usually seen in turbulent fluid, however later we will see that in Magnus glider case the lamniar flow model still works very well under this Re.

        In case 1, with a relatively flat catapult, the Magnus glider goes up and velocity decreases rapidly. Then it reaches the highest point and finally moves with a constant velocity downward. The results are shown in Fig.~\ref{Trajectory}.

                \begin{figure}[!h]
            \centering
            \includegraphics[width = 0.7\textwidth]{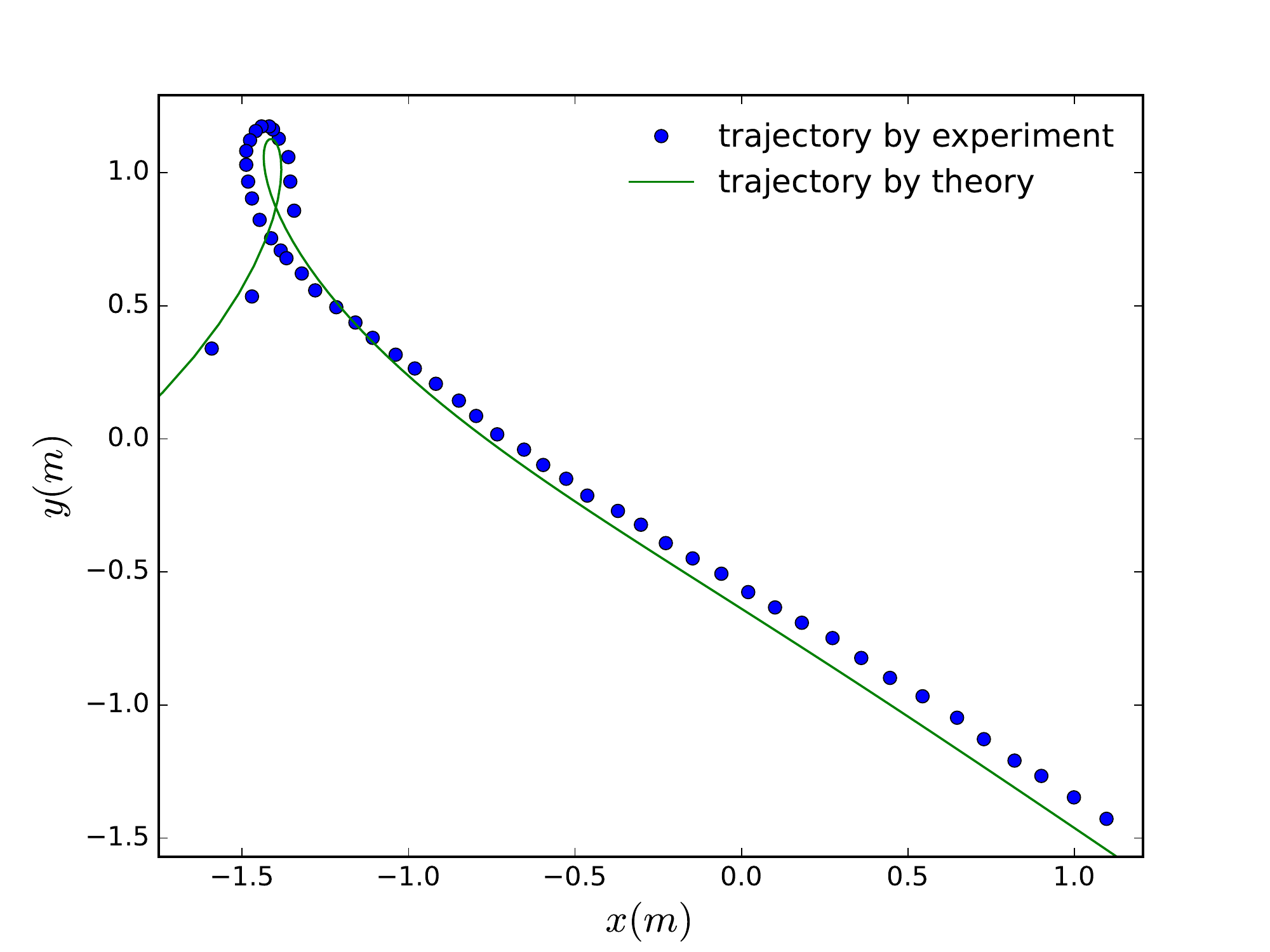}
            \caption{(Case 2) Trajectory of Magnus glider. Green line is the theoretical prediction by our model. The launch angle is ${45^ \circ }$ to the horizontal.}
            \label{Trajectory2}
        \end{figure}

		In the case 2 (Fig.~\ref{Trajectory2}), the geometrical and material parameters of Magnus glider remains the same with the case 1, with $V= 2.823 m/s$ and $\omega = 30.1 \cdot 2\pi ~rad/s$. The main difference with case 1 is that we set the launch angle to be ${45^ \circ }$ up the horizontal. From the result we can see the trajectory of Magnus glider form a loop when it goes up. After that, it remains a constant speed and moved the same as in case 1.

		Generally speaking, the theoretical curve fits well with our experiment result, which shows that our model correctly describes the main parameters which decide the dynamical evolution of Magnus glider. The approximations we made hugely simplify the equations and enable us the ability to make the analytical prediction, while it also leads to a minor deviation of the trajectory in Fig.~\ref{Trajectory}. We ascribe this deviation to following reason: the approximation of our model fail in the area near the y-direction peak of trajectory (mainly in part II in Fig.~\ref{Trajectory}). In part II, the Magnus glider`s velocity $V$ is relatively small, while it still has a considerable rotating speed $\omega$. So for the further study, in this part we could consider modifying the model in part II. Since the deviation is rather small, we can safely say that our result reveals the general laws of the motion of Magnus glider, which is controlled by gravity, Magnus force and the fluid resistance.

\section{Conclusion}

        In this work, we analytically constructed a model for rotating objects in viscous fluid based on the kinematics of Magnus glider, and experimentally studied Magnus force using a wind tunnel. The analytical results show that the Magnus force in this system is proportional to the product of centroid velocity $V$ and angular velocity $\omega$.   The dragging coefficient $\alpha$ is obtained through the wind tunnel experiment. The analytical prediction is consistent with experiments, as $\alpha$ applied well in the prediction of Magnus glider`s actual trajectory. Our work provides a framework for analyzing the generic rotating objects with near-cylinder geometry.

\section{Acknowledgment}

     We sincerely thank Zengming Zhang, Xiuzhe Luo, Lingyuan Ji and Han Chen for valuable discussion. We also thank USTC Centre of Physical Experiments for providing the space for the experiments.

\bibliographystyle{unsrt}
\bibliography{bibtex.bib}

\end{document}